\documentclass[aps,prd,twocolumn,groupedaddress,showpacs]{revtex4}

\usepackage{graphicx}

\begin{document}


\title{Heavy Superheated Droplet Detectors as a Probe of Spin-independent WIMP Dark Matter Existence}


\author{F. Giuliani}
\email[]{franck@cii.fc.ul.pt}
\author{T. Morlat}
\affiliation{Centro de F\'isica Nuclear, Universidade de Lisboa,
1649-003 Lisboa, Portugal}
\author{TA Girard}
\affiliation{Centro de F\'isica Nuclear, Universidade de Lisboa,
1649-003 Lisboa, Portugal}

\collaboration{The SIMPLE collaboration}

\date{\today}

\begin{abstract}
At present, application of Superheated Droplet Detectors (SDDs) in
WIMP dark matter searches has been limited to the spin-dependent
sector, owing to the general use of fluorinated refrigerants which
have high spin sensitivity. Given their recent demonstration of a
significant constraint capability with relatively small exposures
and the relative economy of the technique, we consider the potential
impact of heavy versions of such devices on the spin-independent
sector. Limits obtainable from a $\mathrm{CF_{3}I}$-loaded SDD are
estimated on the basis of the radiopurity levels and backgrounds
already achieved by the SIMPLE and PICASSO experiments. With 34 kgd
exposure, equivalent to the current CDMS, such a device may already
probe to below 10$^{-6}$ pb in the spin-independent cross section.

\end{abstract}


\maketitle

\section{INTRODUCTION}

The direct search for evidence of weakly interacting massive
particle (WIMP) dark matter continues to be among the forefront
efforts of experimental physics. Such searches are traditionally
classified as to whether for spin-independent or spin-dependent WIMP
channels, of which the first has generally attracted the most
attention. The current status of search efforts is defined by a
number of projects, including DAMA/NaI-NAIAD \cite{damasm,naiad05},
CDMS \cite{cdms06SI}, ZEPLIN \cite{ZEPLINI} and EDELWEISS
\cite{edel}. Because of their target nuclei spins, several of these
``spin-independent" devices also provide significant constraints on
the spin-dependent phase space. In fact, this sector is also largely
constrained by the results of DAMA/NaI-NAIAD and CDMS \cite{FGprl}.

Among the other experiments in the spin-dependent sector are two
using superheated droplet detectors (SDDs): SIMPLE
($\mathrm{C_{2}ClF_{5}}$-loaded) \cite{plb2} and PICASSO
($\mathrm{C_{4}F_{10}}$-loaded) \cite{picasso05}, which have
recently demonstrated an ability to achieve competitive results with
significantly reduced measurement exposures. This is partially
because of their high fluorine content, but also the result of their
intrinsic insensitivity to the majority of common backgrounds which
complicate other types of direct search experiments. The impact is
clear if one considers that the current results were obtained with
detectors of 42 and 19.4 g active mass, respectively, vis-a-vis the
recent Kamioka report \cite{CaF} of essentially equivalent results
with a 28 kgd exposure of a 300 g $\mathrm{CaF_{2}}$ scintillator.

Given this performance, together with the relative inexpensiveness
of the technique, the question naturally arises as to whether or not
SDDs might have a similar impact on spin-independent measurements.
Since the cross section scales with the squares of both the mass
number and the WIMP-nucleus reduced mass, exploring the
spin-independent channel of WIMP interactions suggests a detector
composition with nuclei of a significantly higher mass number.
Although several readily available ``heavy" refrigerants exist (eg.
$\mathrm{CF_{3}Br, CF_{3}I, XeF_{6},...}$), the problem of
density-matching the suspension gels in order to achieve a
homogeneous dispersion of the refrigerant without introducing
additional radio-contaminants, together with the belief that the
current impact in the spin-dependent sector derives almost
exclusively from the fluorine content, has discouraged their
development. For this reason, some recent attention has focused on
the development of a gel-free bubble chamber approach \cite{coupp}.
While avoiding the problems of density-matching, this technique
requires a significant extension of the metastability lifetime of
the refrigerant, which is severely degraded by surface nucleations
on the container walls. Recently, the Chicago group has succeeded in
achieving lifetimes of up to several hours \cite{coupp}.

We discuss in Sec. II the SDD in general and the feasibility of
fabricating a $\mathrm{CF_{3}I}$ device. The expected background
contributions to the device operation are analyzed using the current
results from the SIMPLE and PICASSO experiments. Sec. III provides
projections of the results to be expected from a
$\mathrm{CF_{3}I}$-based search which show that its implementation
has the potential to make a significant contribution to the search
activity. Conclusions are drawn in Sec. IV.

\section{A ``Heavy" SDD}

\subsection{Fabrication Considerations}

A SDD is a dispersion (emulsion) of small droplets of superheated
liquid freon fixed in a hydrogenated gel, each droplet of which
functions as a mini-bubble chamber. Current device constructions
rely on density matching of the gel $(\rho\sim$ 1.3 g/cm$^{3}$) with
the refrigerant in order to produce a homogeneous distribution of
droplets in the gel during its setting in the fabrication process.
In the case of ``heavy" refrigerants, $\rho\sim$ 2 g/cm$^{3}$; the
common practise of adding heavy salts such as $\mathrm{CsCl}$ to the
gel in order to raise its density is discouraged for dark matter
search applications by its introduction of radioactive
contaminations.

An alternative approach, at least in principle, is to match in
viscosity rather than density. An estimate of the minimum viscosity
($\eta$) required to trap the droplets during the fabrication
process is obtained by equating the viscous and Archimede's forces:

\begin{equation}\label{visc}
  \eta = \frac{2r^{2}gt(\rho_{b} -\rho_{0})}{9D}\\
\end{equation}

\noindent where $r$ is the average droplet radius, $D$ is the height
of the gel, $t$ is the time for a droplet to fall a distance $D$,
and $\rho_{b} (\rho_{0})$ is the $\mathrm{CF_{3}I}$ (gel) density.
For $t$ = 1 hour (the time required for the setting of the gel
during cooling), $\rho_{b} (\rho_{0})$ = 2 $\times 10^{3}$
kg/m$^{3}$(1.3 $\times 10^{3}$ kg/m$^{3}$), $r$ = 35 $\times$
10$^{-6}$m and $D$ = 5 $\times$ 10$^{-2}$m, this yields 0.13 kg/m/s.
We have recently succeeded to produce a gel matrix, using the
standard SIMPLE ingredients with the addition of agarose to modify
the viscosity rather than density match, as well as shift upwards
the sol-gel transition temperature \cite{agarose06}, with a measured
$\eta$ = 0.17 kg/m/s. This has permitted production of a prototype
$\mathrm{CF_{3}I}$-based SDD with 1-3 times the concentration of the
SIMPLE devices. The Chicago group has similarly succeeded in
developing a SDD prototype of $\mathrm{CF_{3}I}$ with a
polyacrylamide-based gel \cite{collarpc}.

The process results in a homogeneous distribution of micron-sized
$\mathrm{CF_{3}I}$ droplets, and a device insensitive to $\gamma$'s
and $\beta$'s at lower temperatures while sensitive to reactor
neutron irradiations via the induced recoils of $\mathrm{F}$,
$\mathrm{C}$ and $\mathrm{I}$. Since all direct search experiments
rely on the detection of WIMP-induced nuclear recoil events, the
neutron response provides an understanding of the device response to
WIMPS, and of an essential background component.

In the current fabrication protocol however, and unlike the current
$\mathrm{C_{2}ClF_{5}}$ fabrications \cite{plb2}, about 50\% of the
refrigerant dissolves into the gel due to its high solubility in the
high hydrogen bond content matrix, consistent with the solubility of
$\mathrm{CF_{3}I}$ in water (16\% of the gel) and glycerin (78\% of
the gel). Abrupt bubble nucleation of large droplets in the
suspension leads to an unchecked growth of small fractures in the
gel via absorption of the dissolved refrigerant by the bubbles, and
a relatively rapid degrading of the detector performance. There is
also a significant presence of clathrates hydrates at low
temperature, implying that the device cannot be stored at
temperature below 0$^{\circ}$C because clathrates hydrates break
down locally the metastability of the droplets. Although the current
SIMPLE background level would suggest moderately long lifetimes in
the underground site, various techniques to include the use of
gelifying agents not requiring water as a solvent or the use of
others techniques to inhibit the diffusion of the dissolved gas, are
being explored.

\subsection{Background Considerations}

The sensitivity of a $\mathrm{CF_{3}I}$-based SDD dark matter
measurement is defined by the device response. Following the thermal
spike model of Seitz \cite{seitz}, there are two thresholds for
bubble nucleation: (i) the deposited energy must be larger than the
work of formation of a critical embryo ($E_{c}$), and (ii) $E_{c}$
must be deposited within a distance of the order of a critical
radius.

\begin{figure}[h]
  \includegraphics[width=8 cm]{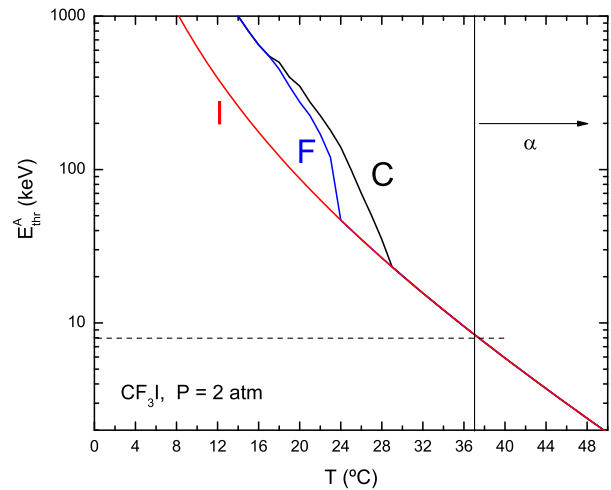}\\
  \caption{Variation of $E_{thr}^{A}$ with temperature for each of
  the $CF_{3}I$ components. The vertical line indicates the temperature
  below which $\alpha$'s are below the stopping power threshold of the detector. }\label{Ethr}
\end{figure}

Fig. \ref{Ethr} shows how the two thresholds combine into the mass
number ($A$)-dependent threshold recoil energy $E_{thr}^{A}$. The
bubble nucleation efficiency of an ion of mass number $A$ recoiling
with energy $E$ is given by the superheat factor
$S_{A}(E)=1-\frac{E_{thr}^{A}}{E}$ \cite{loapf}, set to 8 keV
(37$^{\circ}$C for all the recoiling ions ($\mathrm{C}$,
$\mathrm{F}$ and $\mathrm{I}$) as in the case of the most recent
SIMPLE measurement \cite{plb2}. $E_{thr}^{A}$ can be set as low as
6.5 keV before onset of the gel melting at 40$^{\circ}$C; the
stopping power is $\geq$ 100 keV/$\mu$m for temperatures up to
$\sim$ 5$^{\circ}$C above the gel melting point. In fact, whenever
all the recoiling ions stop within a pressure- and
temperature-dependent critical distance, $E_{thr}^{A}$ and
$S_{A}(E)$ do not depend on A \cite{nima}. $E_{thr}^{A}$ (Fig.
\ref{Ethr}) was calculated as a function of the operating
temperature and pressure by using the method employed by SIMPLE for
R-12- and R-115-loaded SDDs \cite{collar1,collar2,nima}, with
thermodynamic parameters taken from Ref. \cite{Duan} and recoiling
ion stopping powers precalculated with SRIM 2003 \cite{srim}. As
evident, the calculations foresee the same $E_{thr}^{A}$ and $S_{A}$
for $\mathrm{F}$, $\mathrm{C}$, $\mathrm{I}$ at temperatures above
$\sim$ 29$^{\circ}$C (2 atm).

The reason why the threshold recoil energies $E_{thr}^{A}$ for
direct detection of $\mathrm{I}$, $\mathrm{F}$ and $\mathrm{C}$ ions
in Fig. \ref{Ethr} do not coincide for all temperatures is that
while in the range $E_{thr}^{I}$ = $E_{c}$, below $\sim$
24$^{\circ}$C a $\mathrm{F}$ ion above $E_{c}$ and below
$E_{thr}^{F}$ has not enough stopping power to trigger a nucleation.
More generally, a particle above $E_{c}$ but below the stopping
power threshold cannot directly produce a bubble nucleation, and can
only be detected indirectly, with lower efficiency, through a
secondary recoiling ion. Since inelastic scattering will in general
involve the absorption of part of the available energy by either the
recoiling nucleus or the scattered particle, this extends to all
types of scattering.

The energy $E_{total}$ (kinetic+mass) needed by an incident particle
of mass $m$ in order to produce a non-relativistic recoiling ion of
kinetic energy $E_{R}$ and mass number $A$ is given by

\begin{equation}\label{kinema}
\begin{array}{l}
  E_{total}=E_{R}+ \\
  \sqrt{E_{total}^{2}+2AE_{R}-2\sqrt{E_{total}^{2}-m^{2}}\sqrt{2AE_{R}}\cos\theta}\\
\end{array}
\end{equation}

\noindent where $m$ is the incident particle mass, $\theta$ the
recoil angle, and the mass of both nucleons has been approximated to
1 GeV/c$^{2}$. Since $\cos\theta \leq 1$, it is straightforward to
show that $E_{total}$ is at least that corresponding to the
recoiling nucleus linear momentum $\sqrt{2AE_{R}}$:

\begin{equation}\label{blindness}
E_{total} \geq \frac{E_{R} + \sqrt{E_{R}^{2}+4m^{2}+2AE_{R}}}{2}
\geq \frac{\sqrt{2AE_{R}}}{2} .
\end{equation}

\noindent Applying Eq. (\ref{blindness}) to $^{12}\mathrm{C}$, the
lightest isotope in $\mathrm{CF_{3}I}$, it is clear that no light
radiation below $E_{total}=$ 5.4 MeV can produce via elastic
scattering non-relativistic recoiling ions of energy higher than 5
keV. For the chosen threshold of 8 keV, the minimum value of
$E_{total}$ becomes 6.9 MeV. Therefore, under pressure and
temperature operating conditions such that $E_{thr}^{A}$ is 8 keV,
low stopping power light radiations below $E_{total}$ = 6.9 MeV
cannot produce a bubble nucleation either directly or via a
recoiling ion.

Both the $E_{c}$ and $dE/dx$ thresholds can be tuned via the
operating temperature and pressure conditions to render the SDD
insensitive to energetic gamma-rays, X-rays, electrons and other
radiations depositing less than $\sim$ 200 keV $\mu$m$^{-1}$; the
SDD is essentially sensitivity-limited to neutrons and
$\alpha$-particles. This insensitivity is not trivial: the SDD at
37$^{\circ}$C and 2 bar is effectively ``blind" to $\gamma$
backgrounds below 6.9 MeV. Given the $\sim 10^{7}$ evt/kgd
environmental $\gamma$ rate observed in an unshielded 1 kg Ge
detector \cite{gaitskell}, this blindness to $\gamma$'s is
equivalent to an \textit{intrinsic} rejection factor several orders
of magnitude larger than the bolometer experiments with particle
discrimination \cite{newpic2}.

The external background component is primarily muons and
environmental neutrons. At 1500 mwe, the ambient muon flux is $\sim
10^{-6}$ muons/cm$^{2}$s. The response of SDDs, of both low and high
concentrations, to cosmic-ray muons is well-studied
\cite{newpic2,picasso00}: the SDD muon response is similar to that
of $\gamma$'s, with the threshold sensitivity to these backgrounds
occurring for $s = [(T - T_{b} )/(T_{c} - T_{b})] \geq$ 0.5, where
$T_{c}$, $T_{b}$ are the critical and boiling temperatures of the
refrigerant. SDD operation at 37$^{\circ}$C and 2 bar ($s \sim$ 0.3)
is sufficiently below threshold for this contribution to be
neglected \cite{plb2}, and the predominant external background is
neutron.

The response of low concentration SDDs to various neutron fields has
been studied extensively \cite{Errico,harper,loapf}; the high
concentration SDD response to neutrons has been investigated using
sources of $\mathrm{Am/Be}$, $^{252}\mathrm{Cf}$
\cite{newpic2,collar2} and monochromatic low energy neutron beams
\cite{newpic2,nima}, and is in good agreement with thermodynamic
calculations. Since the detectors are further insensitive to neutron
energies below threshold, the neutron contribution can be reduced or
eliminated by external moderation. The background is also addressed
by the gel and the water bath used to maintain the devices at
operating temperature, which are themselves neutron moderators.

Thus, in the case of the SDD, the background issue is almost
entirely determined by the radiopurity of the device construction.
The SDD consists of two components: the refrigerant, and the gel
matrix. The refrigerant determines the response of the device, and
is singly distilled during detector fabrication. The metastability
limit of a superheated liquid is described by homogeneous nucleation
theory \cite{eberhard}, which gives a limit of stability of the
liquid phase at approximately 90\% of the critical temperature for
organic liquids at atmospheric pressure, and an estimate of the
spontaneous nucleation rate of

\begin{equation}\label{sponucl}
R_{s} = N_{p}\sqrt{\frac{2\tau}{\pi
m}}e^{-\frac{16\pi\tau^{3}}{3kT(P_{l}-P_{v})^{2}}} ,
\end{equation}

\noindent where $N_{p} = \frac{N_{A}\rho _{l}}{M_{mol}}$ = 6.15
$\times 10^{21}$ cm$^{-3}$ with $N_{A}$ Avogadro's number, $M_{mol}$
the molar mass and $\rho_{l}$ the liquid density; $\tau$ is the
surface tension, $m$ is the molecular mass, $k$ is Boltzman's
constant, and $P_{l} (P_{v})$ is the liquid (vapor) pressure. At
40$^{\circ}$C, $R_{s}$ $\sim$ $10^{-1800}$ nucleations/kgd, and
decreases by three order of magnitude per degree with decreasing
temperature: at 37$^{\circ}$C this contribution is entirely
negligible, as verified to within experimental uncertainties in Ref.
\cite{picasso00}.

The gel is the key factor in considerations of the device
backgrounds. The current SIMPLE gel ingredients used in the
$\mathrm{CF_{3}I}$ prototype are purified using pre-eluted
ion-exchanging resins specifically suited to actinide removal; the
freon is single distilled; the water, double distilled. The presence
of $\mathrm{U/Th}$ contaminations in the gel, measured at $\leq$ 0.1
ppb via low-level $\alpha$ spectroscopy, yields an overall
$\alpha$-background level of $< 0.5$ evts/kg freon/d. The $\alpha$
response of SDDs has been studied extensively
\cite{newpic2,collar2}. The SRIM-simulated dE/dx for $\alpha$'s in
$\mathrm{CF_{3}I}$ has a Bragg peak at 700 keV and $\sim 193$
keV/$\mu$m, which sets the temperature threshold for direct $\alpha$
detection to $\sim$ 37$^{\circ}$C (Fig. \ref{Ethr}). Below this
temperature, $\alpha$'s can only be detected through
$\alpha$-induced nuclear recoils. Radon contamination is low because
of the 2 atm overpressure, water immersion, and short $\mathrm{Rn}$
diffusion lengths of the SDD construction materials (glass, metal);
the measured $\mathrm{Rn}$ contamination of the glass is at a level
similar to that of the gel.

\section{POTENTIAL IMPACTS}

In order to assess the potential of a ``heavy" SDD to contribute to
the dark matter search, we assume from the above discussion the
background of a potential $\mathrm{CF_{3}I}$-based search to be at
the level reported by the SIMPLE dark matter search \cite{plb2},
which is 1 evt/kgd above an 8 keV recoil threshold. We further
assume, for comparison purposes, a 34 kgd exposure equivalent to
that of CDMS-II (easily achievable, \textit{e.g.}, by operating for
34 days a detector array with only 1 kg active mass). The expected
number of background events is then 34. Assuming 34 events are
``observed", the highest total expected event number compatible at
90\% C.L. with the ``observation" is 42.8, implying an upper limit
to the total rate of 1.26 evts/kgd. Subtracting the 34 expected
background events from this highest total number of expected events,
the 90\% C.L. upper limit on the WIMP rate expected from the
proposed measurement is 0.26 evts/kgd.

Currently, the discrimination capability of the SDD experiment is
limited to the rejection of coincident device signals in the
detector mosaic, which addresses only penetrating neutrons. The
bubble nucleation process is however a four-stage process
\cite{sound}, the last two of which can generate an acoustic pulse
and the last of which generally provides the recorded signal. The
formation of a high temperature, high pressure zone (stage 2) is
followed by its rapid expansion (10$^{-9}$ s) to a size at which the
pressure inside the bubble almost equals the external pressure; if
the bubble diameter is above a critical length, the fourth stage
then sets in. The bubble expansion in the third stage is soley
attributable to the transformed energy of the incident particle,
whereas the fourth stage is due to the energy stored in the liquid.
The full signal should therefore consist of both a fast and slow
pulse, the fast component of which depends on the nature of the
incident radiation \cite{sound}. To what extent this is resolvable
remains in question, and the feasibility of measuring this stage as
a discrimination technique using ultrasound technology is being
explored.

If the experiment were further able to discriminate each of the 34
``observed" events as background, then it could be claimed that no
WIMP has been observed, and the 90\% C.L. upper limit to the WIMP
rate becomes simply $\frac{\ln 10}{34}$=0.068 evts/kgd, obtained by
setting to 10\% the probability of observing no WIMP. This would
also be the limit in the unlikely case of only 26 events, since at
90\% C.L. at least 26 background events should be observed. We also
show this limit, indicated as ``no evts", with the actual result to
be obtained lying somewhere between the two contours.

\subsection{Spin-independent Sector}

\subsubsection{Isospin-independent}

\begin{figure}[h]
  \includegraphics[width=8 cm]{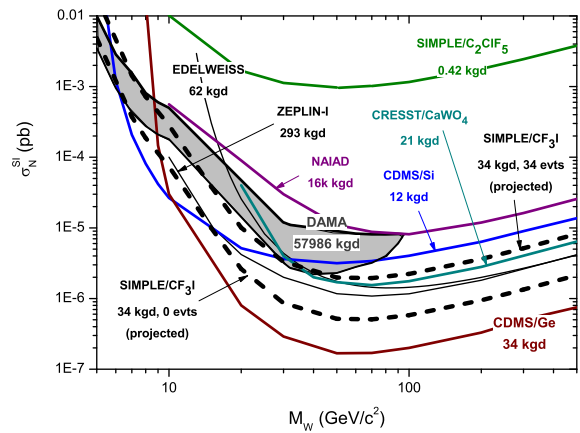}\\
  \caption{Spin-independent exclusion plots for $\mathrm{CF_{3}I}$, projected
  for 34 kgd exposure and 8 keV threshold recoil energy, or 37$^{\circ}$C
  and 2 atm operating temperature and pressure. The lower dashed
  contour assumes background discrimination with no residual events after the
  cut, while the upper dashed line assumes a 34 evts/kgd 90\% C.L. upper
  limit on the undiscriminated WIMP rate (see text).}\label{SIexc}
\end{figure}

Based on the above estimates of the upper limit on the WIMP rate,
and following the procedures of Ref. \cite{Lewin}, the projected
spin-independent limits of Fig. \ref{SIexc} are displayed with those
of some leading experiments
\cite{cdms,ZEPLINI,edel,danai,papCRESSTII}. These projections assume
a standard spherical isothermal halo with a local density of 0.3
GeV/c$^{2}$, a halo velocity of 230 km/s, average Earth velocity of
244 km/s, and a galactic escape velocity of 600 km/s. A Helm form
factor ($F(qr_{n}) = 3\frac{j_{1}(qr_{n})}{qr_{n}} e^{-(qx)^{2}/2}$)
with nuclear radius $r_{n} = \sqrt{z^{2} + \frac{7}{3}\pi^{2}y^{2} -
5x^{2}}$ ($x$ = 1 fm, $y$ = 0.52 fm, $z$[fm] = 1.23$A^{\frac{1}{3}}-
0.6$) is assumed \cite{Lewin}.

As seen in Fig. \ref{SIexc}, with an exposure equivalent to that of
CDMS, an experiment based on $\mathrm{CF_{3}I}$-loaded SDDs would
exclude a significant part of the 3$\sigma$ C.L. DAMA region even
without background discrimination. This is particularly true for the
unexcluded region below $\sim$ 20 GeV, which is better probed by the
$\mathrm{CF_{3}I}$ than CDMS/$\mathrm{Ge}$ owing to the light nuclei
presence. Since the halo velocity distribution has a cutoff at the
galactic escape velocity $v_{esc}$, a nucleus of mass number $A$ in
a detector of threshold $E_{thr}^{A}$ can only reveal WIMPs such
that

\begin{equation}\label{ethre}
    E_{thr}^{A}\leq\frac{2M_{W}^{2}A m_{p}}{(M_{W}+A
m_{p})^{2}}(v_{esc}+v_{E})^{2}   ,
\end{equation}

\noindent where the small difference in neutron and proton mass
\textbf{\textit{($m_{p}$)}}, has been neglected; $v_{E}$ is the
Earth velocity with respect to the Galaxy, so that $v_{esc}$+$v_{E}$
is the maximum WIMP speed with respect to the laboratory. Eq.
(\ref{ethre}) implies a threshold sensitivity in $M_{W}$ given by

\begin{equation}\label{MWthre}
    M_{W min}=\frac{A m_{p}}{\sqrt{\frac{2A
m_{p}}{E_{thr}^{A}}}(v_{esc}+v_{E})-1}   .
\end{equation}

\noindent As $M_{W}$ approaches $M_{Wmin}$, the fraction of the
incident WIMP current detectable through a nucleus of mass number
$A$ vanishes, and, if $A$ is the lightest isotope, the exclusion
plot has a vertical asymptote for $M_{W}=M_{Wmin}$. With the above
parameters, a $\mathrm{CF_{3}I}$ detector with $E_{thr}^{A}$ = 8 keV
can probe down to $\sim$ 3 GeV/c$^{2}$ with $^{12}\mathrm{C}$ and
$\sim$ 3.6 with $^{19}\mathrm{F}$, while a germanium detector with
$E_{thr}^{A}$ = 7 keV is sensitive down to $\sim$ 5.9 GeV/c$^{2}$
through $^{70}\mathrm{Ge}$, and a silicon detector with same
threshold to $\sim$ 3.9 GeV/c$^{2}$ through $^{28}\mathrm{Si}$.

\subsubsection{Isospin-dependent}

Fig. \ref{SIexc} is obtained under either the assumption of
isospin-independence or the approximation that Z $\approx$ N. For
heavy nuclei, the latter breaks down, giving the experiments with a
heavy component a sensitivity to a possible isospin-dependence of
the WIMP interaction. Without \textit{a priori} assuming
isospin-independence, the cross section $\sigma_{A}$ for the
scattering of a spin-independent WIMP on a nucleus of mass number
$A$ is given by \cite{Lewin}

\begin{equation}\label{sigmanuc}
\sigma_{A}=\frac{4}{\pi}G_{F}^{2}\mu_{A}^{2}(g_{p}Z+g_{n}N)^{2}  ,
\end{equation}

\noindent where $G_{F}$ is the Fermi constant, $\mu_{A}$ the
WIMP-nucleus reduced mass and $g_{p,n}$ the WIMP-proton and
WIMP-neutron spin-independent coupling strengths (\textit{i.e.} the
coupling coefficients, in units of $\sqrt{2}G_{F}$).

\begin{figure}[h]
  \includegraphics[width=8 cm]{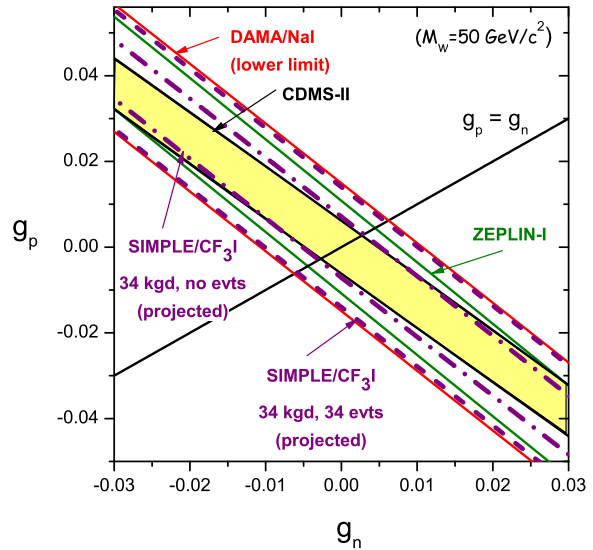}\\
  \caption{Exclusion plots in the $g_{p}-g_{n}$ plane at 50 GeV/c$^{2}$.
  The projections for ``no evt" (thick dash-dot) and ``34 evt" (thick dash)
  34 kgd $\mathrm{CF_{3}I}$ SDD exposures are seen to contribute to the overall
  picture (the overall allowed region is shaded). The ``lower limit" of
  DAMA/$\mathrm{NaI}$ is the inner boundary of its allowed shell.}\label{gg50}
\end{figure}

Fig. \ref{gg50} shows the resulting spin-independent
$\mathrm{CF_{3}I}$ exclusion projection for a 34 kgd exposure at
$M_{W}=$50 GeV/c$^{2}$ following Ref. \cite{SInd}, together with the
current results of some leading searches. In this representation, as
clear from Eq. (\ref{sigmanuc}), the exclusion plots for a given
$M_{W}$ are generally ellipses. Even at this exposure level, a
$\mathrm{CF_{3}I}$ SDD experiment could already contribute to the
overall exclusion.

\subsection{Spin-dependent Sector}

\begin{figure}
  \includegraphics[width=8 cm]{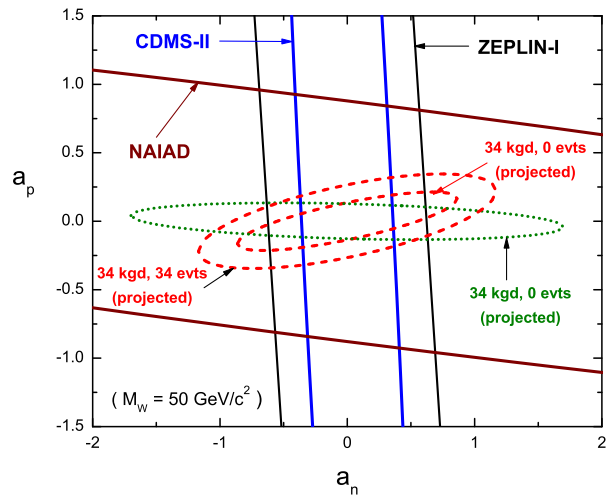}\\
  \caption{Spin-dependent exclusions in the $a_{p}-a_{n}$ plane at $M_{W}=$50
  GeV/c$^{2}$, indicating the anticipated limits from ``no evt" and ``34 evt"
  34 kgd $\mathrm{CF_{3}I}$ measurements, with the spin matrix elements from Ref.
  \cite{Pachestro}; the dotted ellipse is a similar 34 kgd ``no evt"
  projection using the spin matrix elements of Ref. \cite{Divari}.}\label{aasurv}
\end{figure}

Although intended for the spin-independent search effort, the
$\mathrm{CF_{3}I}$ device remains sensitive in the spin-dependent
sector, essentially through $^{19}\mathrm{F}$ and
$^{127}\mathrm{I}$. The spin-dependent exclusions for $M_{W}=$50
GeV/c$^{2}$ are shown in Fig. \ref{aasurv}. As evident, with only 34
kgd the projected experiment would supersede the 16389 kgd NAIAD
\cite{naiad05}, or possibly yield a positive signal at this $M_{W}$.
The 50 GeV/c$^{2}$ is chosen since it lies near the minimum of the
various contours of Fig. \ref{SIexc}; for larger or smaller $M_{W}$,
all results are generally less restrictive, and vary differentially.

The details of the isotopic composition are given in Table
\ref{tabisos}. For completeness, $^{13}\mathrm{C}$ is included: its
spins were evaluated in the odd group approximation \cite{bedny}.

\begin{table}[h]
\caption{\label{tabisos}$\mathrm{CF_{3}I}$ spin parameters.}
\begin{ruledtabular}
\begin{tabular}{|c|c|c|c|c|c|c|}
\hline
  Isotope & Z & J$^{\pi}$ & $\langle S_{p}\rangle$ & $\langle S_{n}\rangle$ & abundance & Ref.\\ \hline
  $^{19}$F & 9 & 1/2$^{+}$ & 0.441 & -0.109 & 100\% & \cite{Pachestro,Tovey}\\ \hline
  $^{19}$F & 9 & 1/2$^{+}$ & 0.4751 & -0.0087 & 100\% & \cite{Divari}\\ \hline
  $^{127}$I & 53 & 5/2$^{+}$ & 0.309 & 0.075 & 100\% & \cite{Ressell}\\ \hline
  $^{12}$C & 6 & 0$^{+}$ & 0 & 0 & 98.9\% & \\ \hline
  $^{13}$C \footnotemark & 6 & 1/2$^{-}$ & 0 & -0.184 & 1.1\% & \\ \hline
\end{tabular}

\footnotetext{Calculated in the odd group approximation using Ref.
\cite{isotable}}
\end{ruledtabular}
\end{table}

Since $^{127}$I recoils can have high linear momentum, the zero
momentum transfer limit does not apply to this isotope, and the
non-zero momentum transfer method of Refs.
\cite{Engel,Ressell,Savage} must be applied in order to evaluate
cross section limits from rate limits. Following a common choice,
the results of Ref. \cite{Ressell} for a Bonn A potential have been
used.

The two projections for $\mathrm{CF_{3}I}$ differ in the choice of
the $^{19}\mathrm{F}$ shell model: the dotted ellipse is based on
spin matrix elements (and structure functions) taken from Ref.
\cite{Divari}, while calculation of the dot-dash ellipse employs the
form factor of Ref. \cite{Lewin}, which is independent of $a_{p,n}$.
Here, the difference in orientation of the two projected ellipses is
explained by the 92\% lower $\langle S_{n}\rangle$ estimate of Ref.
\cite{Divari} with respect to the result of Ref. \cite{Pachestro}.
This leads to evaluate a smaller $^{19}\mathrm{F}$ neutron
sensitivity, which tends to stretch the ellipse in the $a_{n}$
direction, and make it more horizontal. Incidentally, this shows
that the spin-dependent response is fluorine-dominated, as expected
from the 3$\times$ larger amount of fluorine with respect to iodine.

\section{SUMMARY}

At least two groups have succeeded in confronting the
density-matching difficulties of ``heavy" SDD fabrication by
viscosity modification, and demonstrated the feasibility of
producing high concentration prototype devices for R\&D.

Given the fabrication feasibility, the pursuit of such experiments
depends on the results to be obtained with the device
implementation, which are seen to contribute competitively in both
spin-independent and -dependent searches. Because of the tunable
double thermodynamic thresholds of the device, a sensitivity
approaching that of the present bolometric CDMS-II experiment could
be economically achieved with a $\mathrm{CF_{3}I}$-loading and
similar exposure. This competitiveness is largely based on the
device insensitivity to a majority of the backgrounds in the more
traditional search devices; to remain competitive will require
techniques of discriminating the remaining contribution. Since the
recoil threshold can be tuned to as low as 6.5 keV, the results in
particular would address the low mass region of the spin-independent
parameter space still insufficiently explored by such searches. The
same measurement would simultaneously contribute to the search in
the spin-dependent sector, either ruling out the DAMA/NaI result
completely or -- more interestingly -- obtaining a positive signal.

Given that the devices are robust, low maintenance and modular, and
generally inexpensive in both construction and operation, large mass
experiments can be easily envisioned. A small-scale measurement
would provide information on the actual sensitivities of such an
experiment, as well as new limits in the spin-independent sector.

\begin{acknowledgments}
We thank A. Brown for useful discussion of the spin matrix elements
of $^{19}$F. We also thank Juan Collar and Brian Odom for useful
discussions regarding their $\mathrm{CF_{3}I}$ bubble chamber
development. F. Giuliani is supported by fellowship
SFRH/BPD/13995/2003 of the Portuguese Foundation for Science and
Technology (FCT); T. Morlat is supported by POCI grant
FIS/57834/2004, of FCT co-financed by FEDER. The activity is also
supported in part by POCI grants FP/63407/2005 and FIS/56369/2004 of
FCT, co-financed by FEDER.
\end{acknowledgments}


\end{document}